\documentclass[5p,sort&compress]{elsarticle}

\usepackage{amsmath}
\usepackage{amssymb}
\usepackage{bm}
\usepackage{fleqn}
\usepackage{geometry}
\usepackage{natbib}
\usepackage{graphicx}
\usepackage{float}
\usepackage{algorithm}
\usepackage{algorithmicx}
\usepackage{algpseudocode}

\newcommand{\bQ}{\ensuremath{\bar{Q}}}
\newcommand{\Non}{\ensuremath{\mathbb{N}^{\mathrm{on}}}}

\begin{document}

\title{Energy-Minimizing Bit Allocation For Powerline OFDM With
  Multiple Delay Constraints}

%% \author{\IEEEauthorblockN{Kaemmatat Jiravanstit and Wiroonsak Santipach}\\
%% \IEEEauthorblockA{Department of Electrical Engineering\\
%% Faculty of Engineering, Kasetsart University\\
%% Bangkok, 10900 Thailand\\
%% Email: {\tt kaemmatat.j@gmail.com} and {\tt  wiroonsak.s@ku.ac.th}}}
\author{Kaemmatat Jiravanstit}
\ead{kaemmatat.j@gmail.com}
\author{Wiroonsak Santipach\corref{cor1}}
\ead{wiroonsak.s@ku.ac.th}
\cortext[cor1]{Corresponding author}
\address{Department of Electrical Engineering, Faculty of Engineering, Kasetsart
  University, Bangkok, 10900, Thailand}
%
%
%% \thanks{K. Jiravanstit and W. Santipach are with the Department of 
%%   Electrical Engineering; Faculty of Engineering; Kasetsart
%%   University, Bangkok, 10900, Thailand (email: kaemmatat.j@gmail.com;
%%   wiroonsak.s@ku.ac.th).}
%%   }%

%\author{Author 1 and Author 2} 

%\markboth{Submitted to IEEE Wireless Communications Letter, Sep. 2015}{}%
%{Shell \MakeLowercase{\textit{et al.}}: Bare Demo of IEEEtran.cls for Journals}

\begin{abstract}
We propose a bit-allocation scheme for powerline orthogonal
frequency-division multiplexing (OFDM) that minimizes total transmit
energy subject to total-bit and delay constraints.  Multiple delay
requirements stem from different sets of data that a transmitter must
time-multiplex and transmit to a receiver.  The proposed bit
allocation takes into account the channel power-to-noise density ratio
of subchannels as well as statistic of narrowband interference and
impulsive noise that is pervasive in powerline communication (PLC)
channels.  The proposed scheme is optimal with 1 or 2 sets of data,
and is suboptimal with more than 2 sets of data.  However, numerical
examples show that the proposed scheme performs close to the optimum.
Also, it is less computationally complex than the optimal scheme
especially when minimizing total energy over large number of data
sets.  We also compare the proposed scheme with some existing schemes
and find that our scheme requires less total transmit energy when the
number of delay constraints is large.
\end{abstract}

\begin{keyword}
Bit allocation \sep energy minimization \sep delay constraints \sep OFDM \sep powerline channels
\end{keyword}
\maketitle

%%%%%%%%%%%%%%%%%%%%%%%
\section{Introduction}
%%%%%%%%%%%%%%%%%%%%%%%

Existing electrical wiring makes powerline communication (PLC) an
enticing technology for home or in-car networking due to cost
reduction in wire installation and for hospitals due to absence of
radio-frequency transmission that could interfere with medical
instruments.  In IEEE 1901 standard, which stipulates broadband
transmission over powerlines~\cite{homeplug}, orthogonal
frequency-division multiplexing (OFDM) is applied to transmit data
over subchannels.  However, designing and optimizing OFDM transmission
require the PLC channel models that accurately reflect the
characteristics of the actual channels.  The work by~\cite{zim} has
proposed the powerline channel models, which are widely used to
analyze the performance of data transmission over PLC channels.  The
models account for both signal attenuation and multipath propagation
due to impedance mismatching of electrical components.  In addition to
colored background noise, impulsive noise caused by power-supply
switching must also be considered when designing PLC transmission
schemes~\cite{herath2012}.  Whenever impulsive noise occurs, its power
could exceed 50 dB above that of the background noise~\cite{zim02}.
The other potential impediment to PLC transmission is narrowband
interference caused by short-wave or medium-wave radio transmitters
with a frequency range between 300 kHz and 3 MHz~\cite{thokozani2013}.

Optimizing bit, subcarrier, and power allocation for both wired and
wireless OFDM have been considered by many works~\cite{shen1,
  kaiming2009, Chaudhuri, Shin, hamini2011, pancaldi2016, ecti2013,
  costas2005}.  In~\cite{shen1}, subcarrier and power allocation
scheme was proposed to maximize sum rate of all users, subject to
proportional rate constraint.  The work by~\cite{kaiming2009} proposed
an adaptive power loading to minimize the transmit power, subject to a
fixed bit rate and maximum bit error rate (BER).  For PLC channels,
several work~\cite{Chaudhuri, Shin, hamini2011, pancaldi2016,
  ecti2013, costas2005} consider resource allocation at the
transmitter with varying objectives. In~\cite{Chaudhuri,
  pancaldi2016}, throughput is maximized with constraints on total
power or BER while in~\cite{Shin}, outage probability is minimized
with a target BER.  In~\cite{hamini2011}, the maximum transmission
time of all subchannels is minimized for a Zimmermann channel
model~\cite{zim}, subject to a total-energy constraint.
In~\cite{ecti2013}, bit-loading with a single delay constraint is
performed on OFDM channels to minimize transmission energy. The work
by~\cite{costas2005} proposed subchannel and bit allocation for
multiple users to minimize transmission power under transmit rate
constraints.

The aim of this work is to minimize transmit energy in PLC OFDM
channels.  Differing particularly
from~\cite{costas2005,hamini2011,ecti2013} and other existing work,
our proposed bit allocation is applied to a transmitter under multiple
delay constraints.  In our problem formulation, a transmitter must
time-multiplex different sets of data with varying delay and total-bit
constraints and then distributes them over OFDM subchannels.  Each set
of data may arise from different users or classes of service that
share the same channel.  For two delay constraints, we derive the
optimal bit allocation over OFDM subchannels and the optimal transmit durations
that minimize the total transmit energy.  With three or more delay
constraints, we propose a suboptimal allocation that iterates the
optimal 2-delay solution and show via numerical results that the
proposed suboptimal allocation performs close to the optimum.  With
numerical examples, we compare our proposed scheme with that
by~\cite{costas2005} and show that our scheme attains lower transmit
energy when the number of delay constraints and the number of
transmitted bits are large.  The proposed scheme also has much lower
computational complexity than the optimum does especially for a large
number of delay constraints.

%%%%%%%%%%%%%%%%%%%%%%%%%%%%%%%%%%%%%%%%%%%%%%%
\section{System Model and Problem Statement}
%%%%%%%%%%%%%%%%%%%%%%%%%%%%%%%%%%%%%%%%%%%%%%%
\label{sys_mod} 

We assume OFDM with $N$ subchannels and that transmitter has $K$
different sets of data with various sizes and delay requirements to
transmit over time-division multiplexing (TDM) network. For the $k$th
data set on the $n$th subchannel, the discrete-time received symbol is
given by
\begin{equation}
  r_{n,k} = \sqrt{P_{n,k}} H_n x_{n,k} + w_{n,k}
\end{equation}
where $x_{n,k}$ is a transmitted symbol with zero mean and unit
variance, $H_n$ is a complex frequency response of the $n$th
subchannel, $P_{n,k}$ is a transmission power for the $k$th set of
data in the $n$th subchannel, and $w_{n,k}$ is a background noise over
the $n$th subchannel for the $k$th data set.  We assume that channel
response is relatively static during the transmission of the current
data sets and thus, frequency response $H_n$ does not depend on the
index $k$.  In other words, the channel's coherent time is
sufficiently long that optimizing the transmission is meaningful.  The
considered channel might correspond to a large PLC network in which
the dynamics of a few nodes do not alter channel response drastically.

In addition to background noise, PLC channels suffer high-energy
impulsive noise, which is peaky in time domain, and narrowband
interference in some environment.  During an instance of impulsive
noise or narrowband interference, the receiver is not able to reliably
decode the received symbols.  Thus, the associated achievable rate
during that instance is reduced to zero.  We denote the probability of
impulsive noise and/or narrowband interference occurrence in
subchannel $n$ by $p_n$.  While each instance of impulsive noise
affects almost all subchannels, narrowband interference from
short-wave or medium-wave radio only affects a few subchannels. Thus,
$p_n$ may not be the same for all subchannels.

With probability $1-p_n$, the achievable rate in bits per second per
Hertz for subchannel $n$ is given by
\begin{equation}  
  C_{n,k} = \log_2\left(1 + \frac{J_{n,k} |H_n|^2}{t_k B \eta_n}\right)
\label{cap}
\end{equation}
where $B$ is a subchannel spacing, $\eta_n$ is the noise's power
spectral density (PSD) of the $n$th subchannel, $t_k$ denotes a
transmit duration for the $k$th set of data, and $J_{n,k}$ is the
energy used to transmit data for that duration on the $n$th
subchannel. Hence, the average number of bits transmitted via the
$n$th subchannel is given by
\begin{align} 
  \bQ_{n,k} &= E[C_{n,k}] t_k B\\
  &= (1-p_n) t_k B \log_2\left(1 + \frac{J_{n,k}}{t_k B G_n}\right) 
\label{eqq}
\end{align}
where $G_n \triangleq \frac{\eta_n}{|H_n|^2}$ is an inverse of channel
power-to-noise density ratio (CNR) for subchannel $n$.  The energy
required to transmit average $\bQ_{n,k}$ bits can be computed by
\begin{equation} 
  J_{n,k} = (2^{\frac{\bQ_{n,k}}{(1-p_n)t_k B}} - 1) t_k B G_n.
\label{jn} 
\end{equation}
Assuming that the $k$th set of data consists of total average $Q_k$
bits, the sum of all bits for the $k$th set over all OFDM subchannels must
equal or exceed $Q_k$
\begin{equation} 
  \sum_{n=1}^N \bQ_{n,k} \ge Q_k, \quad \text{for} \ 1 \le k \le K,
  \label{eq:bit_req}
\end{equation} 
while the total energy associated with the $k$th transmission is given
by $\sum_{n=1}^N J_{n,k}$.  We note that the transmit duration for the
$k$th set of data must not exceed its delay requirement $T_k > 0$.
Without loss of generality, sets of data are ordered according to the
required delay in an increasing order.  Thus, $T_1 \le T_2 \le \ldots
\le T_K$. With that transmission order, the delay constraint for the
$k$th set can be expressed as follows
\begin{equation}
 \sum_{i=1}^k t_i \le T_k, \quad \text{for} \ 1 \le k \le K.
 \label{eq:delay_req}
\end{equation}
In other words, the cumulative transmission time up to and including
the $k$th transmission must not exceed $T_k$.  In this work, we would
like to allocate average bits $\{\bQ_{n,k}\}$ over $N$ available
subchannels and transmit durations $\{t_k\}$ for all $K$ sets of data
to minimize the total transmit energy
\begin{equation}
  \min_{\{\bQ_{n,k}\}, \{t_k\}} J \triangleq \sum_{k=1}^{K}
  \sum_{n=1}^{N} J_{n,k}
\label{eq:prob}
\end{equation}
subject to total-average-bit and delay constraints
in~\eqref{eq:bit_req} and~\eqref{eq:delay_req}, respectively.

%%%%%%%%%%%%%%%%%%%%%%%%%%%%%%%%%%%%%%%%
\section{Proposed Bit-Allocation Scheme}
%%%%%%%%%%%%%%%%%%%%%%%%%%%%%%%%%%%%%%%%
\label{sec:bitallo}

First, we consider problem~\eqref{eq:prob} with $K=2$.  We remark that
bit allocation with single data set ($K = 1$) that minimizes transmit
energy, has been solved by our previous work~\cite{ecti2013}.  Since
the objective function in~\eqref{eq:prob} is convex with $\bQ_{n,1},\,
\bQ_{n,2} \ge 0, \forall n$, and $t_1,\, t_2 \ge 0$, while both
constraints~\eqref{eq:bit_req} and~\eqref{eq:delay_req} are linear
with those same variables, the solutions that satisfy the Kuhn-Tucker
conditions are optimal.  To derive the Kuhn-Tucker
conditions~\cite{sun1996}, we first form a Lagrangian as follows
\begin{multline}
  \mathcal{L} = \sum_{n = 1}^{N} (J_{n,1} + J_{n,2}) + \beta_1(t_1 -
  T_1) + \beta_2(t_1 +t_2 - T_2) \\ - \lambda_1(\sum_{n = 1}^{N}
  \bQ_{n,1}-{Q_1}) -\lambda_2(\sum_{n = 1}^{N} \bQ_{n,2}-{Q_2})
\label{mlg}
\end{multline}
where $\beta_1$, $\beta_2$, $\lambda_1$, and $\lambda_2$ are Lagrange
multipliers associated with constraints~\eqref{eq:delay_req}
and~\eqref{eq:bit_req}, respectively. The Kuhn-Tucker conditions
associated with the average bits $\bQ_{n,k}$ and total-bit constraints
are given by
\begin{gather} 
   \frac{\partial \mathcal{L}}{\partial \bQ_{n,k}} = 2^{\frac{\bQ_{n,k}}{(1-p_n)t_k B}}\frac{\ln(2)G_n}{(1-p_n)} - \lambda_k \ge 0, \label{kktm_3}\\
   \bQ_{n,k} \ge 0,\\
   \bQ_{n,k}\frac{\partial \mathcal{L}}{\partial \bQ_{n,k}} = 0, \label{kktm_4}
\end{gather}
for $k =1$ and $2$, and $1\le n \le N$, and
\begin{gather} 
   \frac{\partial \mathcal{L}}{\partial \lambda_k} = - \sum_{n = 1}^{N} \bQ_{n,k} + Q_k \le 0, \label{kktm_5}\\
   \lambda_k \ge 0,\\
   \lambda_k\frac{\partial \mathcal{L}}{\partial \lambda_k} = 0, \label{kktm_6}
\end{gather}
for $k =1$ and $2$.

If subchannel $n$ of set $k$ is active or $\bQ_{n,k} > 0$,
equations~\eqref{kktm_3}-\eqref{kktm_4} imply
\begin{equation}
  \lambda_k = 2^{\frac{\bQ_{n,k}}{(1-p_n)t_k B}}\frac{\ln(2)G_n}{(1-p_n)} .
  \label{eql1}
\end{equation}
But, if that subchannel is not used for transmission or $\bQ_{n,k} =
0$,
\begin{equation}
  \lambda_k - 2^{\frac{\bQ_{n,k}}{(1-p_n)t_k
      B}}\frac{\ln(2)G_n}{(1-p_n)} < 0 .
  \label{eql2}
\end{equation}
For the first set of data ($k=1$), we can solve for the optimal bit
allocation from~\eqref{eql1} and~\eqref{eql2} for some transmit
duration $0< t_1 \le T_1$ as follows
\begin{multline}
  \bQ_{n,1} = B t_1 (1-p_n)\\ \times \log_2 \left( 1 + \left(\frac{1-p_n}{G_n
    \ln(2)} \lambda_1 - 1\right)^+ \right)
  \label{eq:Q}
\end{multline}
where a positive-part function $x^+ = \max\{x, 0\}$. Hence, if $G_n
\ln(2)/(1-p_n) < \lambda_1$, the bit allocation in subchannel $n$ is
nonzero or that subchannel is active.  Given $t_1$, the bit allocation
for each subchannel depends on $G_n/(1-p_n)$ and threshold
$\lambda_1$. If the quality of the subchannel is good ($G_n$ is
small), the bit allocation $\bQ_{n,1}$ will be large. However, if the
quality of the subchannel is so poor that $G_n \ln(2)/(1-p_n) <
\lambda_1$, then that subchannel will not be active and will be
allocated zero bits.  Thus, $\lambda_1$ is the threshold that
activates subchannels to transmit the first set of data.  The
threshold $\lambda_1$ can be determined from the total-bit constraint,
which is obtained by summing~\eqref{eq:Q} over all subchannels as
follows
\begin{multline}
  B \sum_{n=1}^N (1-p_n) \log_2 \left( 1 + \left(\frac{1-p_n}{G_n
    \ln(2)} \lambda_1 - 1\right)^+ \right)\\ = \frac{Q_1}{t_1} .
  \label{eq:thes}
\end{multline}
Note that solving for $\lambda_1$ and $\{\bQ_{n,1}\}$ gives water
filling-like solutions.  However, these solutions are not the same as the
classical power allocation that maximizes sum capacity.  We see
from~\eqref{eq:thes} that $\lambda_1$ increases with the transmission
rate $Q_1/t_1$.  Hence, if $Q_1/t_1$ is large enough, all subchannels
can be active.

The threshold $\lambda_1$ can be solved from implicit
equation~\eqref{eq:thes}.  However, if the number of active
subchannels is known, we can explicitly solve for $\lambda_1$
from~\eqref{eq:thes} as follows
\begin{equation}
\log_2(\lambda_1) = \frac{\frac{Q_1}{t_1 B} - \sum_{n\in \Non_1}
  (1-p_n) \log_2\left(\frac{1-p_n}{G_n \ln(2)}\right)}{\sum_{n\in
    \Non_1} 1-p_n}
\label{eq:ld}
\end{equation}
where $\Non_1 \subset \{ 1,2,\dots,N \}$ is the set of active
subchannels transmitting the first set of data.  With~\eqref{eq:ld},
we propose the following iterations in Algorithm~\ref{al:lambda} to
find the optimal $\lambda_1$ or $\lambda_2$.  For each set of data,
the algorithm takes at most $N$ iterations to find the optimal
threshold $\lambda_k$ and the set of active subchannels $\Non_k$.

For a single set of data ($K = 1$), we set input $t_1$ equal to $T_1$
in Algorithm~\ref{al:lambda}, which will find the optimal bit
allocation $\{Q_{n,1}\}$ for all $n$ and the optimal threshold
$\lambda_1$.  In the case that $K=1$ and the probability of impulsive
noise or interference occurring in a subchannel is the same for all
subchannels, Algorithm~\ref{al:lambda} reverts back to the bit
allocation proposed in our previous work~\cite{ecti2013}.  Regarding
computational complexity, the algorithm takes at most $N$ iterations
to find the optimal threshold $\lambda_k$ and the set of active
subchannels $\Non_k$ for each set of data.

\begin{algorithm}
\caption{Finding the optimal thresholds and bit allocation for the
  $k$th data set}
\label{al:lambda}
\begin{algorithmic}[1]
\Require $Q_k$, $t_k$, $B$, $\{G_n\}$, and $\{p_n\}$.
\State Initialize: $\Non_k[0] = \{1,2,...,N\}$ and $i = 0$ 
\Repeat
   \State $i \gets i + 1$
   \State $\Non_k[i] \gets \Non_k[i-1]$
   \State Compute $\lambda_k$ from~\eqref{eq:ld}.
   \For {$n \in \Non_k[i]$}
		\State Compute $\bQ_{n,k}$ from~\eqref{eq:Q}.
		\If{$\bQ_{n,k} = 0$}
		  $\Non_k[i] \gets \Non_k[i] \backslash \{n\}$
		\EndIf
   \EndFor
\Until{$\Non_k[i] = \Non_k[i-1]$}
\State \Return $\lambda_k$ and $\{\bQ_{1,k}, \bQ_{2,k}, \ldots , \bQ_{N,k}\}$
\end{algorithmic}
\end{algorithm}

To find the optimal transmit durations for 2 data sets ($K = 2$), we
derive the corresponding Kuhn-Tucker conditions for $t_1$ and $t_2$ given by
\begin{multline} 
  \frac{\partial \mathcal{L}}{\partial t_k} = \sum_{n = 1}^{N} \bigg[ (2^{\frac{\bQ_{n,k}}{(1-p_n)t_k B}}-1) B G_n \\- 2^{\frac{\bQ_{n,k}}{(1-p_n)t_k B}}\frac{\bQ_{n,k} G_n \ln(2)}{(1-p_n)t_k}\bigg] - \sum_{i = k}^{K} \beta_i \ge 0, \label{kktm_1}
\end{multline}
\vspace{-0.25in}
\begin{gather}
   t_k \ge 0,\\
   t_k\frac{\partial \mathcal{L}}{\partial t_k} = 0, \label{kktm_2}
\end{gather}
for $k = 1$ and $2$.  Since $t_k > 0$, $\partial \mathcal{L}/ \partial
t_k = 0$.  Substitute~\eqref{eql1} in~\eqref{kktm_1} to obtain
\begin{multline}
  \lambda_k \left( \frac{B}{\ln(2)} \sum_{n \in \Non_k} (1 - p_n)
  -\frac{Q_k}{t_k}\right)\\ - B\sum_{n\in\Non_k} G_n - \sum_{i = k}^{K}
  \beta_i = 0.
\end{multline}

The Kuhn-Tucker conditions for Lagrange multipliers $\beta_1$ and
$\beta_2$ are given by
\begin{gather} 
   \frac{\partial \mathcal{L}}{\partial \beta_k} = T_k - \sum_{i = 1}^{k}t_i \ge 0, \label{kktm_7}\\
   \beta_k \ge 0,\\
   \beta_k\frac{\partial \mathcal{L}}{\partial \beta_k} = 0, \label{kktm_8}
\end{gather}
for $k = 1$ and $2$.  Since the second delay constraint is always
tight or $t_2 = T_2 - t_1$, the associated Lagrange multiplier
$\beta_2 > 0$ is given by
\begin{multline}
  \beta_2 = \lambda_2 \left( \frac{B}{\ln(2)} \sum_{m \in \Non_2} (1 -
  p_m) -\frac{Q_2}{t_2}\right)\\ - B\sum_{m\in\Non_2} G_m.
  \label{eq:b2}
\end{multline}
Note that $\beta_2$ is a function of the rate $Q_2/t_2$, $\lambda_2$,
and $\Non_2$.  The last two parameters can be computed by
Algorithm~\ref{al:lambda}.

Similarly, if the delay constraint for the first set is tight or $t_1
= T_1$, then $\beta_1 > 0$. However, if $t_1 < T_1$, then $\beta_1 =
0$ where
\begin{multline}
  \beta_1 = \lambda_1 \left( \frac{B}{\ln(2)} \sum_{n \in \Non_1} (1 -
  p_n) -\frac{Q_1}{t_1}\right)\\ - B\sum_{n\in\Non_1} G_n - \beta_2.
  \label{eq:b1}
\end{multline}
Given $t_1$, $\beta_1$ can be computed by~\eqref{eq:b1} in conjunction
with~\eqref{eq:b2} and Algorithm~\ref{al:lambda}.  If the optimal $t_1
< T_1$, there must exist at least one value of $t_1 \in (0,T_1)$ that
results in $\beta_1 = 0$.  Thus, instead of a brute-force search of
$t_1$, we propose to perform binary search or bisection
method~\cite{burden1985} with a target accuracy $\delta > 0$ set to be
very small.  For each iteration, the interval for $t_1$ is halved
until $|\beta_1| < \delta$.  After $t_1$ is found, we can compute $t_2
= T_2 - t_1$.  Finally, the optimal transmit durations and bit
allocation for the 2 data sets are obtained.  We summarize the steps
in Algorithm~\ref{al:opt}.

If the maximum delay for the first data set has not been reached or
$t_1 < T_1$, then $\beta_1 = 0$ and thus, combining~\eqref{eq:b2}
and~\eqref{eq:b1} eliminates $\beta_2$.  Since \eqref{eq:ld} shows
that $\lambda_k$ is also a function of $Q_k/t_k$, we can show that
$Q_1/t_1 = Q_2/t_2$. This also holds true for $K > 2$.  Therefore, we
conclude that if delay constraints are not too restricting, the data
rates $Q_k/t_k$ obtained from the proposed scheme will be equal for
all users.

\begin{algorithm}
\caption{Optimal transmit duration and bit allocation over
  subchannels for 2 sets of data}
\label{al:opt}
\begin{algorithmic}[1]
\Require $Q_1, Q_2, T_1, T_2$ where $0 < T_1 \le T_2$, $B$, $\{G_n\}$, and $\{p_n\}$.
\State Set accuracy $\delta \ll 1$.
\State $t_1 \gets T_1$
\State Compute $\lambda_1$ and $\{\bQ_{n,1}\}$ with Algorithm~\ref{al:lambda}.
\State $t_2 \gets T_2 - t_1$
\State Compute $\lambda_2$ and $\{\bQ_{n,2}\}$ with Algorithm~\ref{al:lambda}.
\State Compute $\beta_2$ and $\beta_1$ from~\eqref{eq:b2} and \eqref{eq:b1}.
\If{$\beta_1 >0$ and $\beta_2 > 0$}
\State \Return $t_1$, $t_2$, $\{Q_{n,1}\}$, and $\{Q_{n,2}\}$
\Else
\State $t_l \gets 0$ and $t_r \gets T_1$
\Repeat 
\State $t_1 \gets \frac{t_l + t_r}{2}$
\State Compute $\lambda_1$ and $\{\bQ_{n,1}\}$ with Algorithm~\ref{al:lambda}.
\State $t_2 \gets T_2 - t_1$
\State Compute $\lambda_2$ and $\{\bQ_{n,2}\}$ with Algorithm~\ref{al:lambda}.
\State Compute $\beta_2$ and $\beta_1$ from~\eqref{eq:b2} and \eqref{eq:b1}.
\If{$\beta_1 > 0$}
\State $t_l \gets t_1$
\Else 
\State $t_r \gets t_1$
\EndIf
\Until{$|\beta_1| < \delta$ and $\beta_2 > 0$}
\State \Return $t_1$, $t_2$, $\{\bQ_{n,1}\}$, and $\{\bQ_{n,2}\}$
\EndIf
\end{algorithmic}
\end{algorithm}

For problem~\eqref{eq:prob} with the number of delay constraints $K >
2$, the associated Lagrangian is given by
\begin{multline}
  \mathcal{L} = \sum_{k=1}^{K}\sum_{n = 1}^{N} J_{n,k} -\sum_{k =
    1}^{K}\lambda_k(\sum_{n = 1}^{N} \bQ_{n,k}-{Q_k})\\ +\sum_{k =
    1}^{K}\beta_k(\sum_{i=1}^{k} t_i - T_k)
\label{mlgk}
\end{multline}
where $\lambda_k$ and $\beta_k$ denote Lagrange multipliers associated
with constraints~\eqref{eq:bit_req} and~\eqref{eq:delay_req},
respectively. Similar to the problem with $K=2$, the Kuhn-Tucker conditions
can be obtained.  The threshold $\lambda_k$ can be determined from the
total-bit constraint as follows
\begin{multline}
  B \sum_{n=1}^N (1-p_n) \log_2 \left( 1 + \left(\frac{1-p_n}{G_n
    \ln(2)} \lambda_k - 1\right)^+ \right)\\ = \frac{Q_k}{t_k} .
  \label{eq:thesk}
\end{multline}
Given the rate $Q_k/t_k$, we can solve for $\lambda_k$, which is the
water level that determines the set of active subchannels for set $k$
denoted by $\Non_k$.  If a subchannel is active, the average number of
allocated bits for that subchannel is given by
\begin{multline}
  \bQ_{n,k} = B t_k (1-p_n)\\ \times
  \log_2 \left( 1 + \left(\frac{1-p_n}{G_n
    \ln(2)} \lambda_k - 1\right)^+ \right) .
  \label{eq:Qk}
\end{multline}

Since the last delay constraint is always tight or $t_K = T_K -
\sum_{k=1}^{K-1}t_k$, $\beta_K > 0$ and is given by
\begin{multline}
  \beta_K = \lambda_K \left( \frac{B}{\ln(2)} \sum_{m \in \Non_K} (1 -
  p_m) -\frac{Q_K}{t_K}\right)\\ - B\sum_{m\in\Non_K} G_m.
  \label{eq:bK}
\end{multline}
For other data sets $k \ne K$, if the delay constraint of that set is
tight or $\sum_{i=1}^k t_i = T_k$, then $\beta_k > 0$. Otherwise,
$\beta_k = 0$ where
\begin{multline}
  \beta_k = \lambda_k \left( \frac{B}{\ln(2)} \sum_{n \in \Non_k} (1 -
  p_n) -\frac{Q_k}{t_k}\right)\\ - B\sum_{n\in\Non_k} G_n -
  \sum_{i=k+1}^K\beta_i, \quad \forall k \ne K.
  \label{eq:bk}
\end{multline}

Finding the set of optimal transmit durations $\{t_k\}$ involves
solving a nonlinear system with~\eqref{kktm_7}-\eqref{kktm_8},
\eqref{eq:thesk}, and~\eqref{eq:bK}-\eqref{eq:bk} totaling $5K$
equations, and $3K$ unknowns including the sets of Lagrange
multipliers $\{\beta_k\}$ and $\{\lambda_k\}$.  Once the optimal
$\{t_k\}$ is obtained, bit allocation across subchannels for all sets
can be computed by~\eqref{eq:Qk}.  However, solving for $\{t_k\}$ is
exceedingly complex for a very large $K$.

For a simpler allocation scheme with comparable performance, we
propose Algorithm~\ref{a3:opt}, which iteratively applies the optimal
solutions for two delay constraints ($K=2$) derived earlier.  The
proposed scheme is executed as follows. First, we set $\mathsf{Q}_1 =
\sum_{k=1}^{K-1} Q_k$ and $\mathsf{Q}_2 = Q_K$, and $\mathsf{T}_1 =
T_{K-1}$ and $\mathsf{T}_2 = T_K$.  With $\mathsf{Q}_1$,
$\mathsf{Q}_2$, $\mathsf{T}_1$, and $\mathsf{T}_2$, we apply
Algorithm~\ref{al:opt}, which returns $\{\bar{\mathsf{Q}}_{n,1}\}$,
$\{\bar{\mathsf{Q}}_{n,2}\}$, $\forall n$, $\mathsf{t}_1$, and
$\mathsf{t}_2$.  We obtain the transmit duration for the last or the
$K$th data set, $t_K = \mathsf{t}_2$, and the bit allocation for the
$K$th set on subchannel $n$, $\bQ_{n,K} = \mathsf{Q}_{n,2}$ for $1 \le
n \le N$.  To find the allocation for the $(K-1)$th set, we apply
Algorithm~\ref{al:opt} with $\mathsf{Q}_1 = \sum_{k=1}^{K-2} Q_k$ and
$\mathsf{Q}_2 = Q_{K-1}$, and $\mathsf{T}_1 = T_{K-2}$ and
$\mathsf{T}_2 = T_{K-1}$.  We iterate these steps for other $K-2$
rounds or until all bit allocation and transmit durations are
obtained.  All steps are shown in Algorithm~\ref{a3:opt}.  The
computational complexity of this proposed suboptimal scheme increases
only linearly with $K$ and is much less than that of solving for the
optimum.  Moreover, numerical results will show that these suboptimal
solutions perform close to the optimum.

\begin{algorithm}
\caption{The proposed scheme that finds transmit durations and bit
  allocation over subchannels for the number of delay constraints $K \ge 1$.}
\label{a3:opt}
\begin{algorithmic}[1]
\Require $\{Q_k\}$, $\{T_k\}$ where $0 < T_1 \le T_2 \le...\le T_K$, $B$, $\{G_n\}$, and $\{p_n\}$.
\If{$K > 1$}
\Repeat 
\State $\mathsf{T}_1 = T_{K-1}$
\State $\mathsf{T}_2 = T_K$
\State $\mathsf{Q}_1 = \sum_{k=1}^{K-1}Q_k$
\State $\mathsf{Q}_2 = Q_K$
\State Compute $\mathsf{t}_1$, $\mathsf{t}_2$, $\{\bar{\mathsf{Q}}_{n,1}\}$, and $\{\bar{\mathsf{Q}}_{n,2}\}$ by Algorithm~\ref{al:opt} with inputs $\mathsf{T}_1$, $\mathsf{T}_2$, $\mathsf{Q}_1$, and $\mathsf{Q}_2$, and channel parameters $B$, $\{G_n\}$, and $\{p_n\}$.
\State $t_K \leftarrow \mathsf{t}_2$
\State $\{\bQ_{n,K}\} \leftarrow \{\bar{\mathsf{Q}}_{n,2}\}$ for all $n$
\State $K \leftarrow K-1$
\Until{$K = 1$}
\EndIf
\State $t_1 \gets T_1$
\State Compute $\{\bQ_{n,1}\}$ by Algorithm~\ref{al:lambda} with $Q_1$, and channel parameters $B$, $\{G_n\}$, and $\{p_n\}$.
\State \Return $t_1,t_2, \ldots, t_K$, and $\{\bQ_{n,1}\}, \{\bQ_{n,2}\}, \ldots, \{\bQ_{n,K}\}$
\end{algorithmic}
\end{algorithm}

%%%%%%%%%%%%%%%%%%%%%%%%%%%
\section{Numerical Results}
%%%%%%%%%%%%%%%%%%%%%%%%%%%
\label{num_re}

To generate numerical results, we utilize a frequency response of the
15-path channel model from~\cite{zim}, which was obtained by measuring
and approximating from the actual powerline network.  The measured
network consists of a 110-meter link installed in an estate of
terraced houses with six 15-meter branches.  A link between two points
in the network consists of a distributor cable or a series connection
of distributor cables.  The signal propagated along the line-of-sight
path and non line-of-sight paths as echoes.  Cable loss due to the
length of signal propagation and frequency is also captured in the
model.

For PSD of the background noise, the worst case
in~\cite[Fig. 2]{dibert2011} is assumed.  Typically, the probability
of impulsive noise is less than 5\% even in a highly disturbed PLC
network~\cite{zim02}.  Assuming the worst-case probability, $p_n$ is
at least equal to 0.05. Additionally, subchannels in a frequency range
of short-wave or medium-wave radio can be interfered by those radio
transmitters.  Thus, we assume $p_n = 0.06$ in those subchannels
accounting for interference.  Models and parameters used in the
proceeding numerical results are summarized in
Table~\ref{tb_parameter}.

\begin{table}
\caption{Models and parameters used in the numerical results}
\centering
\begin{tabular}{p{1.75in}c}
\hline
Item & Value \\
\hline
Frequency response & 15-path model~\cite{zim}\\
PSD of colored noise & \cite[Fig. 2]{dibert2011}\\
Frequency range & 0.5-20 MHz\\
Subcarrier spacing ($B$) & 24.414 kHz\\
Number of subchannels ($N$) & 735\\
Maximum transmission rate & 200 Mbps\\
Probability of impulsive noise and interference ($p_n$) &
$\left\{\begin{array}{l} 0.06, 1 \le n \le 39\\ 0.05, 40 \le n \le 735
\end{array}\right.$\\
\hline %inserts single line
\end{tabular}
\label{tb_parameter} % is used to refer this table in the text  
\end{table}

\begin{figure*}
  \centering 
  \includegraphics[width=4.5in]{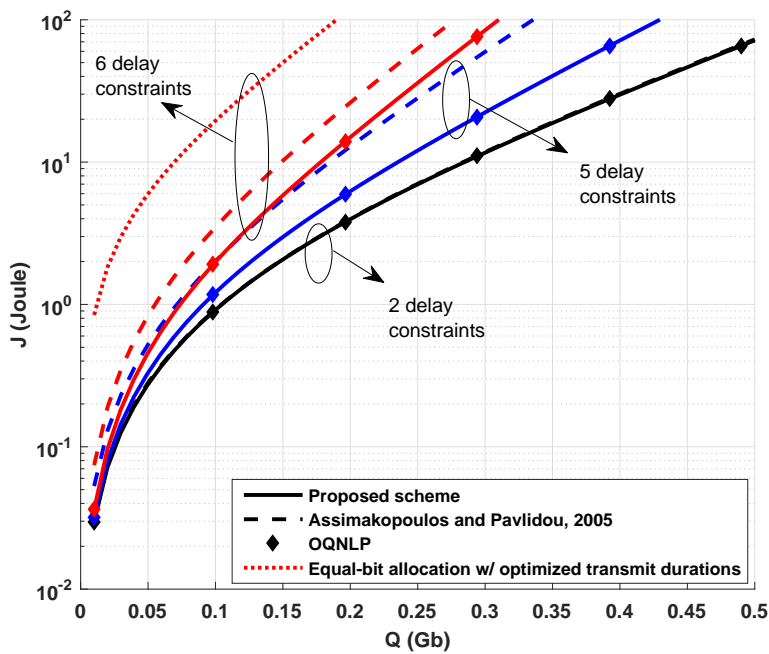}
  \caption{Total transmission energy $J$ from our scheme, that
    by~\cite{costas2005}, and the optimum by OQNLP are shown with
    total average bits $Q$.}
\label{fig_fivedelays} 
\end{figure*}

Fig.~\ref{fig_fivedelays} shows total transmission energy $J$ with
total average $Q$ bits. First, we assume 2 data sets ($K=2$) with $T_1
= T_2 = 5$ s (40 TDM slots), and $Q_1 = 0.25 Q$, and $Q_2 = 0.75
Q$. We remark that delay constraint $T_k$ is set to be a multiple of
TDM time-slot (each TDM time-slot is 125 ms) and $Q_k$ is set such
that the peak transmission rate is not to exceed 200
Mbps~\cite{homeplug}.  The total energy resulted from the proposed
allocation in Algorithm~\ref{al:opt} is shown by a solid line with
that from OptQuest nonlinear program (OQNLP) shown by diamond-shape
markers.  OQNLP confirms that the proposed solutions satisfy the
Kuhn-Tucker conditions and thus, are optimal.  We also apply the
proposed allocation with $K=5$, $T_k = 3.75$ s, $\forall k$, and $Q_k$
from $\{0.12 Q, 0.1 Q, 0.22 Q, 0.16 Q, 0.4 Q\}$.  Although our
proposed allocation is suboptimal when $K > 2$, it performs almost
the same as the optimal solutions obtained by OQNLP.  The same
observation can also be made for the shown results with $K=6$ with
$T_k = 2.5$ s, $\forall k$, and $Q_k$ from $\{0.24Q, 0.1Q, 0.08Q,
0.16Q, 0.2Q, 0.22Q\}$.  We note that as expected, larger energy is
required with larger number of delay constraints.  To demonstrate the
potential energy saving from the proposed scheme, we compare it with
equal-bit allocation in which all subchannels are allocated equal
number of bits regardless of subchannel quality.  The dotted line
shows the total transmit energy for the $K=6$ case with equal-bit
allocation.  By applying the proposed scheme, we observe energy
decrease of almost an order of magnitude for this channel.

For performance comparison, we look for existing schemes with the
similar objective and constraints as ours in the literature.  We have
found a comparable bit-loading scheme in orthogonal frequency-division
multiple access (OFDMA) by~\cite{costas2005} whose objective is to
minimize total transmit power. In~\cite{costas2005}, a transmitter
assigns non-overlapping subchannels to users with different rate
requirements by first allocating the number of subchannels and then
assigning set of subchannels for each user.  For comparison, we set
the same delay requirement for all data sets or users. Total
transmission energy with the allocation scheme by~\cite{costas2005} is
also shown in Fig.~\ref{fig_fivedelays} with dashed lines.  We see
that with 2 data sets or users, our scheme and that
by~\cite{costas2005} perform about the same.  However, with more data
sets and larger amount of data to transmit, our proposed scheme
clearly attains lower total energy than the existing scheme does.  In
an effort to minimize complexity, \cite{costas2005} does not jointly
optimize the number of subchannels and the set of subchannels for each
user.  Hence, there is  some performance degradation.

\begin{figure*}[htb] 
  \centering 
  \includegraphics[width=4.5in]{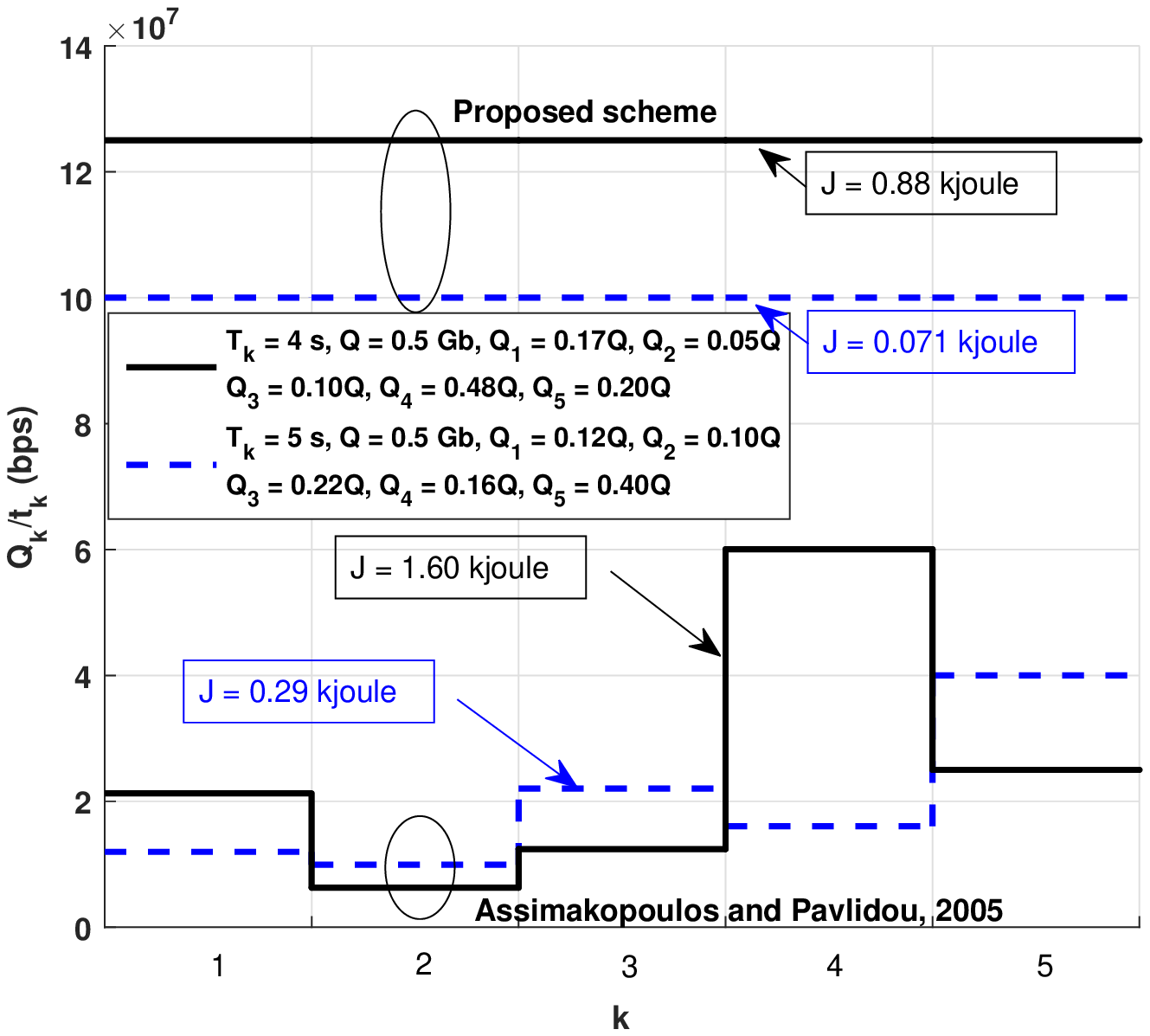}
  \caption{Data rates for all sets/users from the proposed scheme are
    compared with those from the scheme by~\cite{costas2005}.}
\label{fig_rate} 
\end{figure*}

In Fig.~\ref{fig_rate}, we show the data rates resulting from our
bit-allocation scheme for all 5 data sets or users ($k =1, 2, \ldots,
5$).  In the first trial, we set the maximum delay of all sets $T_k =
4$ s and total bits for each set to be $0.17 Q$, $0.05Q$, $0.10Q$,
$0.48Q$, and $0.20Q$, respectively where $Q = 500$ Mb.  We see that
the resulting data rates for all sets are equal to 125 Mbps.  Since
the delay constraints in this trial are not too limiting, our scheme
produces the solution with equal rates as discussed in
Section~\ref{sec:bitallo}.  We compare with the data rates obtained
from~\cite{costas2005} and see that data rate for each user is not the
same.  The data rate from~\cite{costas2005} is simply obtained by
dividing the total bits for a user by the maximum delay. Thus, user 4
with the largest number of bits to transmit has the maximum data
rate. We also display the total transmit energy for both schemes and
see that our scheme consumes close to half of what~\cite{costas2005}
does.  The results from the second trial with different maximum delay
and set of total bits are also shown with dashed lines.  The same
observation for the first trial can also been made.

\begin{figure*}[htb] 
  \centering 
  \includegraphics[width=4.5in]{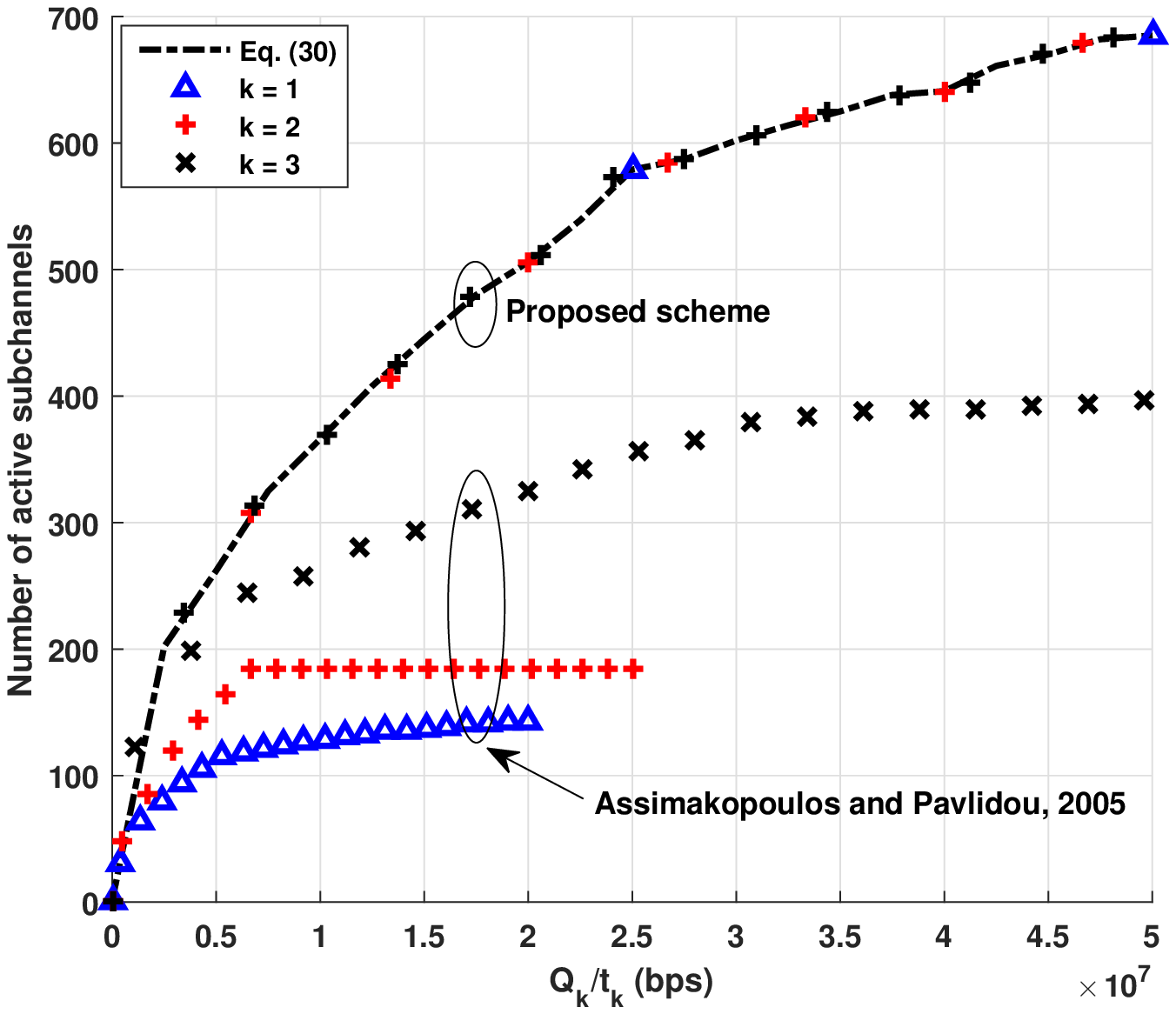}
  \caption{The number of active subchannels is shown with data rate
    for $K=3$.}
\label{fig_non} 
\end{figure*}

In Fig~\ref{fig_non}, we examine the number of active subchannels as
data rate increases.  For our proposed allocation, the number of
active subchannels for data set $k$ depends on the threshold
$\lambda_k$ and rate $Q_k/t_k$ as shown in~\eqref{eq:thesk}. The
dash-dot curve is obtained from solving $\lambda_k$
in~\eqref{eq:thesk} for the given rate and then using~\eqref{eq:Qk} to
find the number of active subchannels.  For the other plots from the
proposed scheme, we set $K=3$ with $T_k = 0.25$, $1$, and $5$ s and
$Q_k = 0.25Q$, $0.2Q$, and $0.55Q$ with increasing $Q$.  The number of
active subchannels for all three data sets is displayed with
different markers and are shown to be on the dash-dot curve as
expected.  We also show the number of active subchannels obtained from
the allocation scheme by~\cite{costas2005} with $T_k = 5$ s, $\forall
k$. We see that with~\cite{costas2005}, the number of active
subchannels for a different data set does not follow the dash-dot
curve. In~\cite{costas2005}, allocating the number of subchannels for
each set will depend on the rates of all sets as well as subchannel
quality.  Since user 3 ($k=3$) has the largest rate requirement of all
three users, the user is allocated the most number of active
subchannels.

\begin{figure*}[htb] 
  \centering 
  \includegraphics[width=4.5in]{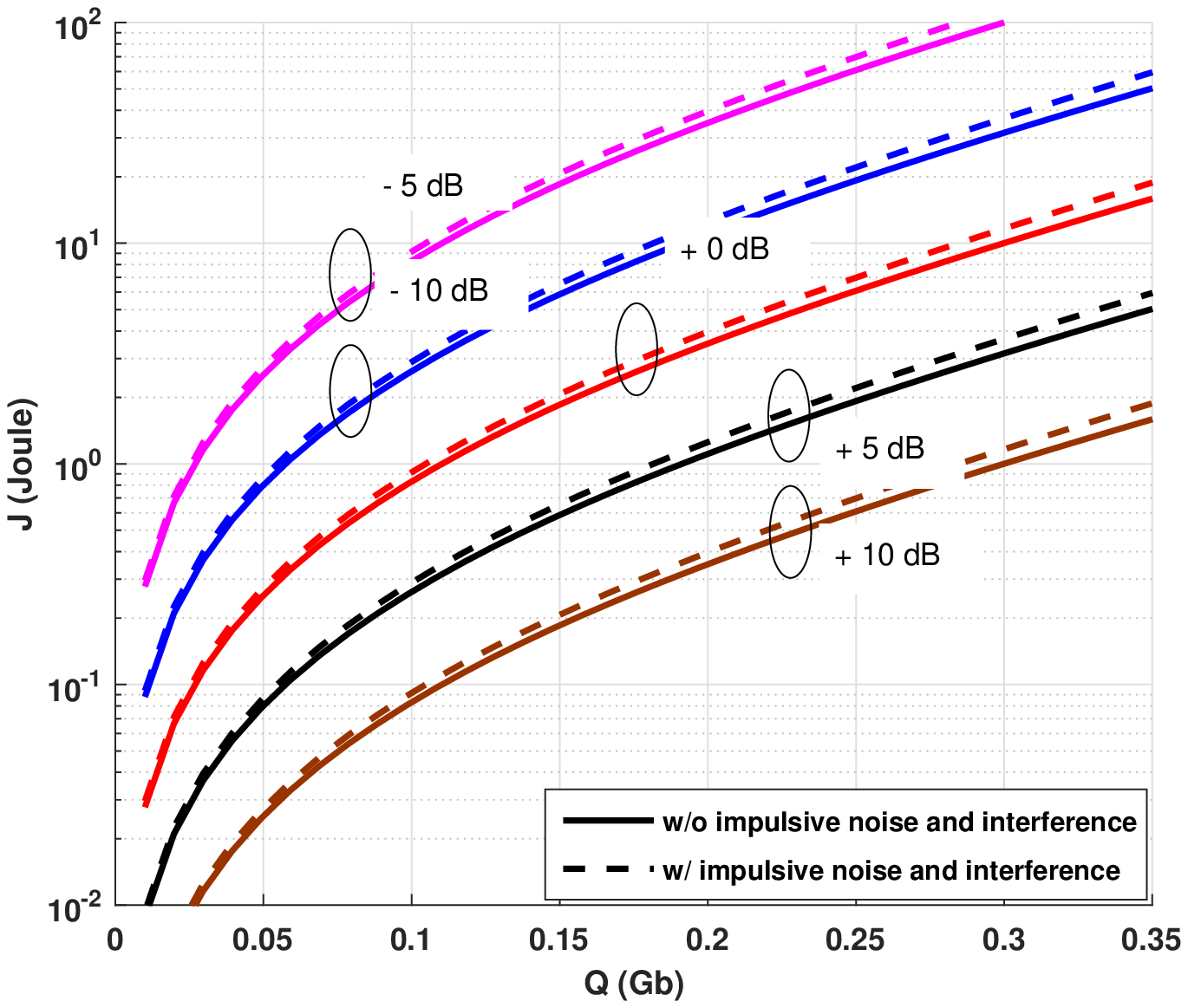}
  \caption{Total transmission energy of the proposed allocation for
    systems with impulsive noise and narrowband interference as
    indicated in Table~\ref{tb_parameter} is compared with that of the 
    ideal systems without impulsive noise and interference.}
\label{fig_varyp} 
\end{figure*}

In Fig.~\ref{fig_varyp}, we show the total energy obtained from the
proposed allocation for systems with $K = 5$, $T_k = 0.625$, $1.25$,
$1.875$, $3.125$, and $5$ s, and the following set of total bits
$\{0.25 Q, 0.10 Q, 0.20 Q, 0.25 Q, 0.10 Q\}$.  We see that the total
transmit energy must increase with the sum of total bits for all data
sets denoted by $Q$.  We also consider the ideal channels with no
impulsive noise or narrowband interference.  It is not surprising that
the total transmit energy is reduced with the ideal channels.
However, we note that energy reduction is small when the number of
total bits $Q$ is small and is increasing with $Q$.  Thus, impulsive
noise and narrowband interference have a larger impact when large number
of bits needs to be transmitted.  We also vary the quality of the
channel or CNR by increasing or decreasing the channel power of all
subchannels by 5 or 10 dB and observe that the energy difference is
large when the average CNR is increased or decreased significantly.

%%%%%%%%%%%%%%%%%%%%%
\section{Conclusions}
%%%%%%%%%%%%%%%%%%%%%
\label{conclude} 

From the study, we find that the optimal bit allocation over
subchannels, the optimal transmit duration, and the number of active
subchannels, depend largely on the transmission rate of each data set
and quality of subchannels, which is indicated by CNR and the
probability of narrowband interference and impulsive noise.  Our
proposed bit allocation is optimal with 2 delay constraints and
performs close to the optimum with more than 2 delay constraints. It
is also less complex than solving for the optimal allocation.  From
numerical results, our scheme is shown to outperform the allocation
scheme by~\cite{costas2005} especially with large amount of data and
larger number of delay constraints.  Systems with narrowband
interference and impulsive noise require higher transmission energy.
Furthermore, the energy increase over that of systems with no or very
small probability of impulsive noise is more pronounced with large
total-bit requirements.  Besides PLC, the proposed bit allocation can
be applied to other applicable OFDM channels such as wireless
multipath channels or wired transmission over twisted-pair cables with
narrowband interference.

%%%%%%%%%%%%%%%%%%%%%%%%%%
\section*{Acknowledgments}
%%%%%%%%%%%%%%%%%%%%%%%%%%

This work was supported by Kasetsart University Research and
Development Institute (KURDI) under the FY2016 Kasetsart University
research grant, and the Royal Golden Jubilee Ph.D. program.

%%%%%%%%%%%%%%%%%%%%%%%%%%%
\section*{References}
%%%%%%%%%%%%%%%%%%%%%%%%%%%

\bibliographystyle{elsarticle-num}
\bibliography{IEEEabrv,plc}
\end{document}